\documentstyle[prd,aps,floats,psfig,twocolumn]{revtex}

\newcommand{\mybib}[2]{\bibitem{#2}}

\newcommand{\ApJ}{Astrophys. J.}
\newcommand{\AJ}{Astron. J.}
\newcommand{\PRL}{Phys. Rev. Lett.}
\newcommand{\PRD}{Phys. Rev. D}
\newcommand{\MNRAS}{MNRAS}
\newcommand{\AsAs}{A\&A}
\newcommand{\aut}[2]{{#2.\ #1}}
\newcommand{\refs}[6]{#2, {\bf #3} {#4} (#5)}

\newcommand{\amp}{and }

\def\sun{\hbox{$\odot$}}

\long\def\comment#1{}

\def\la{\hbox{ \raise.35ex\rlap{$<$}\lower.6ex\hbox{$\sim$}\ }}
\def\ga{\hbox{ \raise.35ex\rlap{$>$}\lower.6ex\hbox{$\sim$}\ }}

\def\W2{{\cal W}}

\newcommand{\deld}{\delta^{\rm D}}
\newcommand{\bn}{\hat{\bf n}}
\newcommand{\bm}{\hat{\bf m}}

\newcommand{\bk}{\hat{\bf k}}
\newcommand{\veck}{{\bf k}}
\newcommand{\vecl}{{\bf l}}
\newcommand{\rad}{r}  
\newcommand{\da}{d_A} 

\newcommand{\isw}{{\rm ISW}}
\newcommand{\X}{{\rm X}}

\newcommand{\sky}{{\rm sky}}

\newcommand{\Ylmn}{Y_{l}^{m}}
\newcommand{\alm}[1]{a_{l_#1 m_#1}}

\begin{document}
\twocolumn[\hsize\textwidth\columnwidth\hsize\csname
@twocolumnfalse\endcsname

\title{The integrated Sachs-Wolfe Effect -- Large Scale Structure Correlation}
\author{Asantha Cooray}
\address{
Theoretical Astrophysics, California Institute of Technology,
 Pasadena, California 91125.\\
E-mail: asante@caltech.edu}

\date{Submitted to PRD}

\maketitle

\begin{abstract}
We discuss the correlation between late-time integrated Sachs-Wolfe (ISW)
effect in the cosmic microwave background (CMB) temperature anisotropies and
the large scale structure of the local universe. 
This correlation has been proposed 
and studied in the literature as a probe of the dark energy and its
physical properties. We consider a variety of large scale structure tracers 
suitable for a detection of the ISW effect via a 
cross-correlation. In addition to luminous
sources, we suggest the use of tracers such as dark matter halos
or galaxy clusters.
A suitable  catalog of mass selected halos for this purpose 
can be constructed with upcoming wide-field lensing and 
Sunyaev-Zel'dovich (SZ) effect surveys. With multifrequency data, the presence
of the ISW-large scale structure correlation can also be investigated through
a cross-correlation of the frequency cleaned SZ and CMB maps.
While convergence maps constructed from lensing surveys of the
large scale structure via galaxy ellipticities are less 
correlated with the ISW effect,
lensing potentials that deflect CMB photons are strongly 
correlated and allow, probably, the best
 mechanism to study the ISW-large scale structure correlation with 
CMB data alone. 
\end{abstract}
\vskip 0.5truecm



]

\section{Introduction}

It is by now well known the importance of 
cosmic microwave background (CMB) temperature fluctuations as a probe
of cosmology \cite{est}. In addition to the dominant anisotropy 
contribution at the 
last scattering surface \cite{Huetal97}, CMB photons, while on
transit to us, encounter the  large scale structure which imprints
 modifications on the temperature fluctuations.
In general, large scale structure affects CMB through two distinct processes:
gravity and Compton scattering. The modifications due to gravity arises
from frequency changes via gravitational red and blue-shifts 
\cite{SacWol67,ReeSci68}, through
deflections involving lensing \cite{Blaetal87,Hu00} and
time-delays \cite{HuCoo01}. During the reionized epoch,
photos can both generate  and erase primary fluctuations
through scattering via free electrons \cite{SunZel80,OstVis86}.

Here, we discuss an effect due to gravitational redshift
commonly known in the literature as the
integrated Sachs-Wolfe (ISW; \cite{SacWol67}) effect at late 
times. The temperature fluctuations in the ISW effect result from the 
differential redshift effect from photons climbing in and out of time
evolving potential perturbations from last scattering surface to
the present day.  In currently popular 
cold dark matter cosmologies with a cosmological constant,
significant contributions arise  at redshifts of cosmological constant
domination ($z \lesssim 2$), at 
on and above the scale of the horizon at the time of decay.
When projected on the sky, 
the ISW effect contributes at large angular scales and has a power
spectrum that scales with the wavenumber 
as $k^{-5}$ times the linear density field power 
spectrum \cite{HuSug94}. This is in contrast to
most other contributions to CMB temperature fluctuations
from the local universe, such
as the well known thermal Sunyaev-Zel'dovich (SZ; \cite{SunZel80}) effect, 
that peak at small angular scales and scales with the wavenumber as $k^{-1}$. 

Since time evolving potentials that contribute to the ISW effect may also
be probed by observations of the large scale structure, it is
then expected that the ISW effect may be correlated with
certain tracers. The presence of the ISW effect can then be
 detected via a cross-correlation
of the CMB temperature fluctuations at large angular scales and
the fluctuations in an appropriate tracer field.
Since the ISW contribution is sensitive to how one models
cosmology at late-times, such as the presence of a dark 
energy component and its physical properties (for example, 
the ratio of dark energy pressure to density),
the correlation between the CMB temperature  and tracer fields
has been widely discussed in the literature \cite{CriTur96}.

Though several attempts have already been made to 
cross-correlate the ISW effect, using the best 
COBE temperature map, and foreground 
sources, such as X-ray and radio galaxies, there is, so far,
 no clear detection of the correlation signal \cite{Bouetal98}. 
As we discuss later, these non-detections are not surprising given
the large sample variance associated with the correlating part of the 
temperature fluctuations, i.e., the ISW effect, with
contribution to the variance coming from the primary temperature
fluctuations. Even for the best case scenarios involving whole-sky
observations and no noise contribution in the tracer field, 
the expected  cumulative signal-to-noise for the ISW-large scale 
structure correlation is at most a ten. If the ISW-large
scale structure correlation is to be used as probe of cosmological
and astrophysical properties, it is certainly necessary to
study in detail  what tracers are best correlated with the
ISW effect and why.

In this paper, we address this question by studying
in detail the correlation between the ISW effect 
and large scale structure. Since we are primarily interested in
understanding what types of tracers are best suited to detect the
correlation,  we will consider a variety of
large scale structure observations and tracers. 
These include luminous sources such as galaxies or active galactic
nuclei (AGN) at different wavelengths, 
the dark matter halos or galaxy clusters that describe the 
large scale clustering of the universe, gravitational lensing,
and other contributions to CMB from the large scale structure, such
as the thermal SZ effect.
Unlike previous studies \cite{PeiSpe00,Veretal01}, we do not consider
cosmological applications of the cross-correlation involving
measurement of dark matter and dark energy properties.

In \S~\ref{sec:method}, we outline the correlation between ISW effect and
large scale structure.
In \S~\ref{sec:discussion}, we discuss our results and study which tracer 
is best suited for
correlation studies. We suggest that in addition to tracers
involving sources such as galaxies
or AGNs,  dark matter halos  that contain these
sources may also be  
suitable for a detection of the ISW effect via a cross-correlation.
The best probe of the ISW-large scale structure
correlation is the lensing effect on CMB photons
on transit to us from the last scattering surface. Since one can
extract information related to lensing deflections from quadratic statistics
on temperature data \cite{SelZal99,Hu01}, 
this allows one to use all-sky CMB anisotropy maps, 
such as from the Planck surveyor, 
alone to extract the ISW-large scale structure correlation.

\section{ISW-Large Scale Structure Correlation}
\label{sec:method}

The integrated Sachs-Wolfe effect \cite{SacWol67} results from
the late time decay of gravitational potential fluctuations. The
resulting
temperature fluctuations in the CMB can be written as
\begin{equation}
T^\isw(\bn) = -2 \int_0^{\rad_0} d\rad \dot{\Phi}(\rad,\bn \rad) \, ,
\end{equation}
where the overdot represent the derivative with respect to conformal
distance, or equivalently look-back time, from the
observer at  redshift $z=0$
\begin{equation}
\rad(z) = \int_0^z {dz' \over H(z')} \,.
\end{equation}
Here, the expansion rate for adiabatic CDM cosmological models with a
cosmological constant is
\begin{equation}
H^2 = H_0^2 \left[ \Omega_m(1+z)^3 + \Omega_K (1+z)^2
              +\Omega_\Lambda \right]\,,
\end{equation}
where $H_0$ can be written as the inverse
Hubble distance today $H_0^{-1} = 2997.9h^{-1} $Mpc.
We follow the conventions that 
in units of the critical density $3H_0^2/8\pi G$,
the contribution of each component is denoted $\Omega_i$,
$i=c$ for the CDM, $g$ for the baryons, $\Lambda$ for the cosmological
constant. We also define the 
auxiliary quantities $\Omega_m=\Omega_c+\Omega_b$ and
$\Omega_K=1-\sum_i \Omega_i$, which represent the matter density and
the contribution of spatial curvature to the expansion rate
respectively.

Writing multipole moments of the temperature fluctuation field
$T(\hat{\bf n})$,
\begin{equation}
a_{lm} = \int d\bn T(\bn) \Ylmn {}^*(\bn)\,,
\end{equation}
we can formulate the angular power spectrum as
\begin{eqnarray}
\langle \alm{1}^* \alm{2}\rangle = \deld_{l_1 l_2} \deld_{m_1 m_2}
        C_{l_1}\,.
\end{eqnarray}
For the ISW effect, multipole moments are
\begin{eqnarray}
a^{\rm ISW}_{lm} &=&i^l \int \frac{d^3\veck}{2 \pi^2}
\int d\rad   \dot{\Phi}(\veck) I_l(k)  \Ylmn(\hat{\veck}) \, ,
\nonumber\\
\label{eqn:moments}
\end{eqnarray}
with $I_l^\isw(k) = \int d\rad W^\isw(k,\rad) j_l(k\rad)$, and the window
function for the ISW effect, $W^\isw$.
Here, we have used the Rayleigh expansion of a plane wave
\begin{equation}
e^{i{\bf k}\cdot \hat{\bf n}\rad}=
4\pi\sum_{lm}i^lj_l(k\rad)Y_l^{m \ast}(\bk)
\Ylmn(\bn)\,.
\label{eqn:Rayleigh}
\end{equation}

In order to calculate the power spectrum involving the
time-derivative of potential fluctuations, we make use of the
cosmological Poisson equation \cite{Bar80}.
In Fourier space, we can relate fluctuations in the potential to
the density field as:
\begin{equation}
\Phi = {3 \over 2} \frac{\Omega_m}{a} \left({H_0 \over k}\right)^2
        \left( 1 +3{H_0^2\over k^2}\Omega_K \right)^{-2}
        \delta(k,\rad)\,.
\label{eqn:Poisson}
\end{equation}
Thus, the derivative of the potential can be related to a derivative
of the density field and the scale factor $a$. 
In linear theory, the density field may be scaled backwards to a higher 
redshift by the use of the growth function $G(z)$, where
$\delta(k,r)=G(r)\delta(k,0)$ \cite{Pee80}
\begin{equation}
G(r) \propto {H(r) \over H_0} \int_{z(r)}^\infty dz' (1+z') \left(
{H_0
\over H(z')} \right)^3\,.
\end{equation}
Note that in the matter dominated epoch $G \propto a=(1+z)^{-1}$.
It is, therefore, convenient to define a new variable $F(\rad)$ such that
$F \equiv G/a$.

Since we will be discussing the correlation between the ISW effect
and the tracers of the large scale structure density field, for simplicity,
we can write the power spectrum of ISW effect
in terms of the density field such that
\begin{equation}
C_l^\isw = {2 \over \pi} \int k^2 dk P_{\delta\delta}(k)
                \left[I_l^\isw(k)\right]^2 \,,
\label{eqn:clexact}
\end{equation}
where
\begin{equation}
I_l^\isw(k) = \int_0^{\rad_0} d\rad W^\isw(k,\rad) j_l(k\rad) \, ,
\end{equation}
with
\begin{equation}
W^\isw(k,\rad) = -3\Omega \left(\frac{H_0}{k}\right)^2 \dot{F} \, .
\end{equation}
Here, we have introduced the power spectrum of density fluctuations
\begin{equation}
\left< \delta({\bf k})\delta({\bf k')} \right> = (2\pi)^3
\deld({\bf k}+{\bf k'}) P(k)\,,
\end{equation}
where
\begin{equation}
\frac{k^3P_{\delta \delta}(k)}{2\pi^2} = \delta_H^2
\left({k \over H_0} \right)^{n+3}T^2(k) \,,
\end{equation}
in linear perturbation theory.
Here, $\delta_H$ is the amplitude of the present-day density fluctuations
at the Hubble scale and
we use the fitting formulae of \cite{EisHu99}
in evaluating the transfer function $T(k)$ for CDM models.

The cross-correlation between the ISW effect and another tracer of the
large scale structure can be similarly constructed by taking the multipole
expansion of the tracer field. Following arguments similar to the above, we 
obtain this cross-correlation as
\begin{equation}
C_l^{\isw-\X} = {2 \over \pi} \int k^2 dk P_{\delta\delta}(k)
                I_l^\isw(k) I_l^\X(k) \,,
\end{equation}
where
\begin{equation}
I_l^\X(k) = \int_0^{\rad_0} d\rad W^\X(k,\rad) j_l(k\rad) \, ,
\end{equation}
with the window function for the $X$-field as
$W^\X(k,\rad)$. 
To estimate how well the tracer field correlates with the ISW effect,
we will introduce the correlation coefficient $r_{\isw - \X}$ such that
\begin{equation}
r_{\isw-\X} = \frac{C_l^{\isw-\X}}{\sqrt{C_l^\isw C_l^\X}} \, ,
\end{equation}
where the power spectrum of the X-field follows
\begin{equation}
C_l^{\X} = {2 \over \pi} \int k^2 dk P_{\delta\delta}(k)
                \left[I_l^\X(k)\right]^2 \, .
\end{equation}
A correlation coefficient of $\sim 1$ suggests that the tracer field is
well correlated with the ISW effect. One of the goals of this
paper is to investigate what tracer fields lead to correlation coefficients
close to 1.
 
Using the covariance of the cross-correlation power spectrum, we can write
the signal-to-noise for the detection of the ISW-large scale structure
as
\begin{eqnarray}
&& \left(\frac{\rm S}{\rm N}\right)^2 = f_{\rm sky}  
\sum_{l=l_{\rm min}}^{l_{\rm max}}
(2l+1)  \nonumber \\
&\times& 
\frac{\left[C_l^{\isw-\X}\right]^2}{\left[C_l^{\isw-X}\right]^2 +
\left(C_l^\isw+C_l^{\rm N_\isw}\right)\left(C_l^{\X}+C_l^{\rm N_\X}\right)} \, ,
\label{eqn:sn}
\end{eqnarray}
where noise contributions to the ISW map and the tracer map
are given by $C_l^{\rm N_\isw}$ and $C_l^{\rm N_\X}$, respectively.

In the case of the ISW effect, the noise contributions are effectively
\begin{equation}
C_l^{\rm N_\isw} = C_l^{\rm CMB}+C_l^{\rm det} \, ,
\end{equation}
where $C_l^{\rm CMB}$ is the total temperature  fluctuation contribution,
including the ISW contribution, while $C_l^{\rm det}$ is any detector
noise contribution. As we find later, it is  the $C_l^{\rm CMB}$ that
limits the signal-to-noise for the detection of the ISW-large scale
structure correlation.  Therefore, most of the results we present
here apply equally well to both  MAP and
Planck surveyor missions.  We will also discuss
tracer fields, such as the SZ effect, involving information
extracted using temperature data from these satellites. 
In such cases, when the signal-to-noise is different,
we will separately quote them for MAP and Planck.

In equation~\ref{eqn:sn}, we have approximated the number of independent 
modes available for the measurement
of the cross-correlation as $f_{\rm sky} (2l+1)$. The summation
generally starts from $l_{\rm min}=2$, though, when observations
are limited to a fraction of the sky, the low multipoles are
not independent and there is a substantial reduction in the
number of independent modes. To account for the
loss of low multipole modes, one can approximate the summation
such that $l_{\rm min} \sim \pi/\theta_{\rm sky}$ where $\theta_{\rm sky}$
is the survey size in smallest dimension. Since the ISW effect peaks
at large angular scales, where in some cases the cross-correlation is
also, it is crucial that the reduction in signal-to-noise due to
partial sky coverage be considered. 

Since we will be primarily 
interested in all-sky temperature maps from upcoming satellite
missions, these issues are of minor concern. We will assume a usable
fraction of $f_{\rm sky}=0.65$ for such missions due to sky cuts involving
galactic plane etc, which still leads to almost independent modes at
each multipole. When cross-correlating with surveys of limited 
area tracer fields, however, the detectability 
of the ISW effect via the attempted correlation may crucially depend on
the sky coverage of the tracer field.

Note that an expression of the type in
equation~(\ref{eqn:clexact}) can be evaluated
efficiently with the Limber approximation \cite{Lim54}.
Here, we employ a version based on
the completeness relation of spherical Bessel functions
\begin{equation}
\int dk k^2 F(k) j_l(kr) j_l(kr')  \approx {\pi \over 2} \da^{-2} \deld(r-r')
                                                F(k)\big|_{k={l\over d_A}}\,,
\label{eqn:ovlimber}
\end{equation}
where the assumption is that $F(k)$ is a slowly-varying function.
Using this, we obtain a useful approximation for the
power spectrum for sufficiently high $l$ values, $l \sim$ few hundred, as
\begin{equation}
C_l^\isw = \int d\rad \frac{\left[W^\isw\right]^2}{\da^2}
                P_{\dot{\Phi}\dot{\Phi}}\left[k=\frac{l}{\da},\rad \right]                \, .
\label{eqn:cllimber}
\end{equation}
Here, the comoving angular diameter distance, in terms of the radial distance,
 is
\begin{equation}
\da = H_0^{-1} \Omega_K^{-1/2} \sinh (H_0 \Omega_K^{1/2} \rad)\,.
\end{equation}
Note that as $\Omega_K \rightarrow 0$, $\da \rightarrow \rad$
and we define $\rad(z=\infty)=\rad_0$.

Although we maintained generality in all derivations, we
now illustrate our results with the currently favored $\Lambda$CDM
cosmological model. The parameters for this model
are $\Omega_c=0.30$, $\Omega_b=0.05$, $\Omega_\Lambda=0.65$ and $h=0.65$.
With $n=1$, we adopt the COBE normalization
for $\delta_H$ \cite{BunWhi97} of $4.2 \times 10^{-5}$,
such that mass fluctuations on the $8 h$ Mpc$^{-1}$
scale is  $\sigma_8=0.86$, consistent with observations on the
abundance of galaxy clusters \cite{ViaLid99}.

\section{Results}
\label{sec:discussion}

\begin{figure}
\centerline{\psfig{file=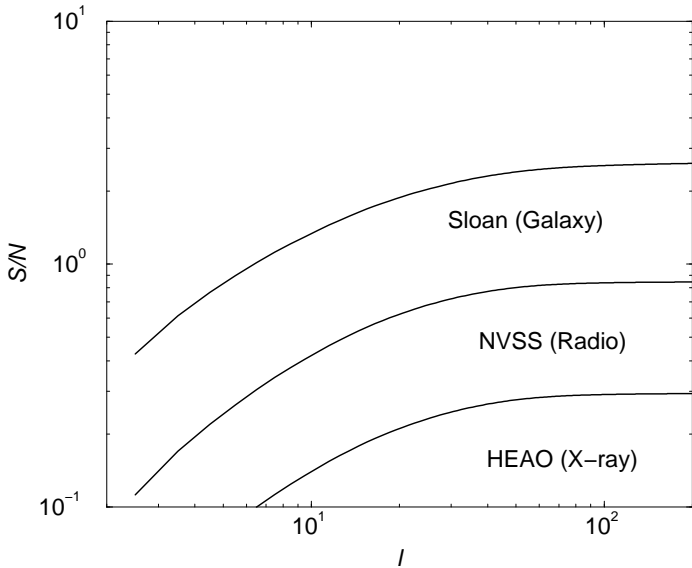,width=3.6in,angle=-90}}
\caption{The signal-to-noise for the detection of source-ISW
cross-correlation with a variety of sources that trace the
large scale structure. The considered surveys include:
HEAO X-ray catalog, radio sources from the NVSS,
and the upcoming Sloan galaxy catalog. The signal-to-noise includes
the fraction of sky covered in each of these surveys.
The correlation with Sloan is limited by the sample variance
associated with the ISW signal while the other two catalogs
have significant shot-noise associated with the limited number counts.}
\label{fig:galsn}
\end{figure}

\begin{figure}
\centerline{\psfig{file=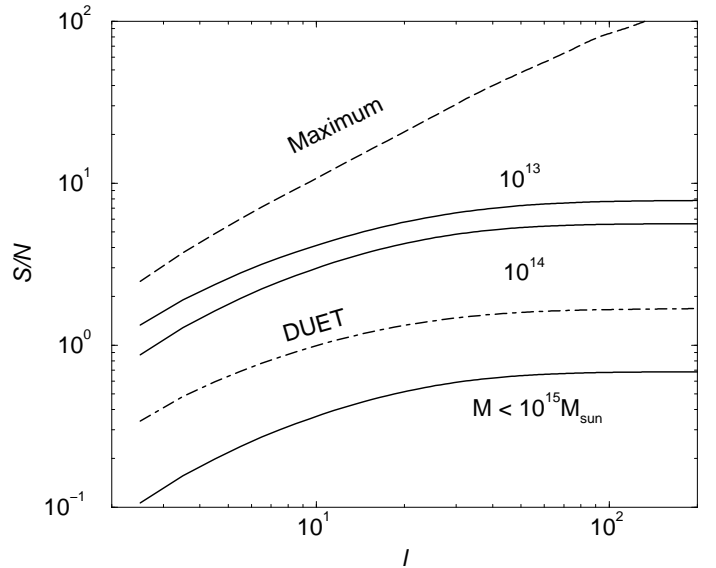,width=3.6in,angle=-90}}
\caption{The signal-to-noise for the detection of source-ISW
cross-correlation with dark matter halos that trace the
large scale structure with low mass limits down to 10$^{15}$, 10$^{14}$
and 10$^{13}$ M$_{\sun}$ (solid lines). 
The long-dashed line is the maximum signal-to-noise when
the ISW effect can be perfectly separated from the dominant CMB data.
The dot-dashed line shows an estimate for the correlation using
a catalog of clusters based on the proposed Dark Universe Exploration Telescope (DUET).}
\label{fig:halosn}
\end{figure}

We now discuss a variety of large scale structure tracers
that can potentially be used to cross-correlate with CMB temperature
data. 

\subsection{ISW-source Correlation}

Probably the most obvious tracer of the
large scale structure density field in the linear regime is luminous sources
such as galaxies at optical wavelengths and AGNs at X-rays and/or radio
wavelengths. We can write the associated window function in this
case as
\begin{equation}
W^\X(k,\rad) = b_\X(k,\rad) n_\X(\rad) G(\rad)\, ,
\end{equation}
where $b_X(k,\rad)$ is the scale-dependent source bias as a function of
the radial distance and $n_X(\rad)$ is the normalized redshift distribution
of sources such that $\int_0^{\rad_0} n_\X(\rad)=1$. 
We will model the redshift distribution in current 
and upcoming source catalogs with an analytic
form:
\begin{equation}
n_\X(z) = \left(\frac{z}{z_0}\right)^{\alpha} \exp -\left(\frac{z}{z_0}\right)^\beta \, ,
\end{equation}
where $(\alpha,\beta)$ denote the slope of the distribution at
low and high $z$'s, respectively with a mean given by $\sim z_0$.
For the purpose of this calculation we take $\alpha=\beta=1$ and
vary $z_0$ from 0.1 to 1.5 so as to mimic the expected
sources from current and upcoming catalogs as well as to
cover the redshift range in which ISW contributions are generally expected.
Since any redshift dependence on bias can
be included as a variation to the source redshift distribution, 
the scale dependence on bias is more important.

Since the ISW effect is primarily associated with large linear scales,
the source bias can be well approximated as scale independent.
Such an assumption is fully consistent with results
from numerical simulations \cite{Benetal00}, results 
from redshift surveys  \cite{Veretal01}
and semi-analytic calculations involving 
the so-called halo model \cite{Sel00}.
We take the source bias to be in the range of 1 to 3 as expected for
galaxies and biased sources as X-ray objects. 
Since $C_l^{\isw-\X} \propto b$,
and $C_l^\X \propto b^2$, note that 
the correlation coefficient is independent of bias
and other normalizing factors. The bias becomes only important
for estimating the signal-to-noise.

In order to estimate signal-to-noise, we describe the noise
contribution associated with source catalogs 
by the finite number of sources
one can effectively use to cross-correlate with temperature data.
We can write this shot-noise contribution as
\begin{equation}
C_l^{\rm N_\X} = \frac{1}{\bar{N}}
\end{equation}
where $\bar{N}$ is the surface density of source per steradian.

In figure~\ref{fig:galsn}, we show cumulative signal-to-noise for
a cross-correlation of the temperature data with large scale structure.
We have considered three source catalogs here involving the HEAO
X-ray catalog ($f_sky=1/3$,$\bar{N} \sim 10^{3}$ sr$^{-1}$ and $b_\X=3$), 
NVSS radio sources ($f_sky=0.82$,$\bar{N}\sim 2 \times 10^{5}$ sr$^{-1}$
and $b_X=1.6$) \cite{Conetal98} and 
galaxy counts down to R band magnitude of 25 from the Sloan Digital
Sky survey ($f_\sky=0.25$, $7 \times 10^{8}$ sr$^{-1}$ and $b_\X=1$)
\cite{Yoretal00}.
The  first two catalogs have already been
used for this cross-correlation with the COBE data \cite{Bouetal98}. 
The non-detection of the cross-correlation is not 
surprising given that we find
cumulative signal-to-noise values less than 1;
this estimate will become worse 
once the noise contribution associated with COBE
temperature data is also included. In both cases, the shot-noise associated
with tracer fields is significant. For Sloan, the limiting
factor in the signal-to-noise is the large noise contribution
associated with the ISW contribution due to dominant primary
temperature fluctuations. Our estimates for the signal-to-noise
for ISW-Sloan correlation is consistent with previous estimates 
\cite{PeiSpe00},
where it was concluded that the detection may be challenging given
the small signal-to-noise.

In addition to luminous sources, one can also cross-correlate the
temperature map with a catalog of galaxy clusters,
or dark matter halos. It is expected that
wide-field galaxy lensing surveys and small-angular 
resolution SZ surveys will allow mass selected catalogs of
clusters \cite{Holetal00}. Similar catalogs of clusters can
also be compiled through wide-field galaxy catalogs such as Sloan and
X-ray imaging data, such as from the proposed Dark Universe Exploration 
Telescope
(DUET).

The redshift  distribution of halos in such
catalogs follows simply from analytical arguments, such as through
the mass function $dn(M,z)/dM$ calculated following analytical
methods such as the Press-Schechter 
(PS; \cite{PreSch74}) mass function or numerical
measurements \cite{Jenetal01}. 
Additionally, the halo
bias is also well known through analytical methods \cite{Moetal97}:
\begin{equation}
b_h(M,z) = 1+ \frac{\left[\nu^2(M,z) - 1\right]}{\delta_c}\, ,
\end{equation}
where $\nu(M,z) = \delta_c/\sigma(M,z)$ is the peak-height threshold.
$\sigma(M,z)$ is the rms fluctuation within a top-hat filter at the
virial radius corresponding to mass $M$, and $\delta_c$ is the
threshold overdensity of spherical collapse. Useful fitting functions
and additional information on these quantities could be found in
\cite{Hen00}. 
For the purpose of this calculation, we use $b_X = \langle b_M \rangle$,
such that the mass averaged halo bias, as a function of
redshift is:
\begin{equation}
\left< b_M \right>(z) = \frac{1}{\bar{n}_h(z)}
      \int_{M_{\rm min}}^{\infty} dM \frac{dn(M,z)}{dM}
b_h(M,z) \, .
\end{equation}
Here, the mean number density of halos, as a function of
redshift, given  by $\bar{n}_h(z) = \int dM
dn(M,z)/dM$. 

The redshift distribution of halos follow similarly
from PS theory: 
\begin{equation}
n_\X(z) = \frac{1}{\bar{N}} \frac{d^2V}{dzd\Omega} 
\left[ \int_{M_{\rm min}}^{\infty} 
dM \frac{dn(M,z)}{dM} \right] \, ,
\end{equation}
where
\begin{equation}
\bar{N} = \int dz \frac{d^2V}{dzd\Omega} 
\left[ \int_{M_{\rm min}}^{\infty} 
dM \frac{dn(M,z)}{dM} \right] \, ,
\end{equation}
with the comoving volume element given by
$d^2V/dzd\Omega$. Note that 
the shot-noise associated with the halo-halo power spectrum is
given by the surface-density of halos $C_l^{\rm N_\X} \equiv 1/\bar{N}$.
Note that even though the surface density of halos are
significantly smaller than, say, the case for galaxies or sources,
one gains an equivalent factor with the increase in
halo bias relative to the same for galaxies. Therefore, on
average, we expect catalogs based on halos to produce same
order of magnitude signal-to-noise as for luminous sources.

In figure~\ref{fig:halosn}, we show the cumulative signal-to-noise for
detecting the ISW-halo correlation. Here, we assume all-sky catalogs of
clusters down to a minimum mass limit of 10$^{13}$, 
10$^{14}$ and 10$^{15}$ M$_{\sun}$.
The lower limit of 10$^{14}$ M$_{\sun}$  is consistent with the
expected mass threshold for SZ clusters that will be detected
with the planned South Pole Telescope \cite{Holetal00}. 
This mass limit is constant over a wide range in redshift, while
mass catalogs based on lensing surveys contain a redshift dependent
minimum mass limit due to the variation in the lensing window function.

The decrease in signal-to-noise with increase in mass is due to
the decrease in the surface density of halos  and thus the increase
in the shot-noise associated with the halo side of the correlation.
Though the halo surface density decreases as a function of
mass, this is partly accounted by the increase in the halo bias
such that even with a low surface density of halos, one can
attempt a correlation of the temperature with cluster catalogs.
As an example of a temperature-source catalog correlation that can
be expected with the MAP data, we also show an estimate for the signal-to-noise
for a catalog of $\sim$ 18000 clusters that is expected to be compiled
from the wide-field X-ray survey by the proposed DUET mission (dot-dashed
line). Though the cumulative signal-to-noise is below 2, the DUET catalog
is significantly preferred over the Sloan survey due to the
fact that cluster bias associated with DUET tracers can be a priori known
through mass estimates based on the electron temperature data while galaxy
bias may be more complicated. 

\begin{figure*}
\centerline{\psfig{file=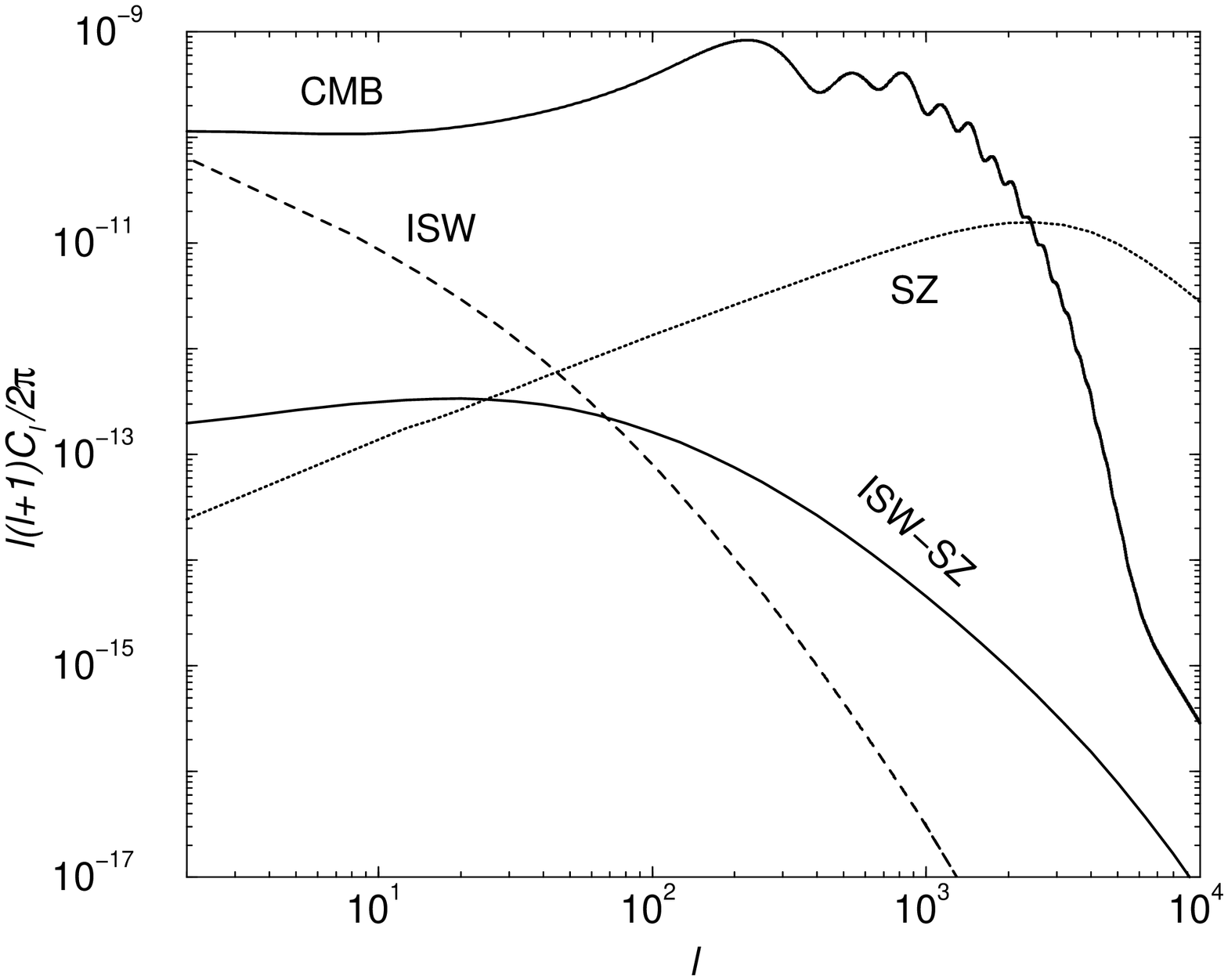,width=3.6in,angle=-90}
\psfig{file=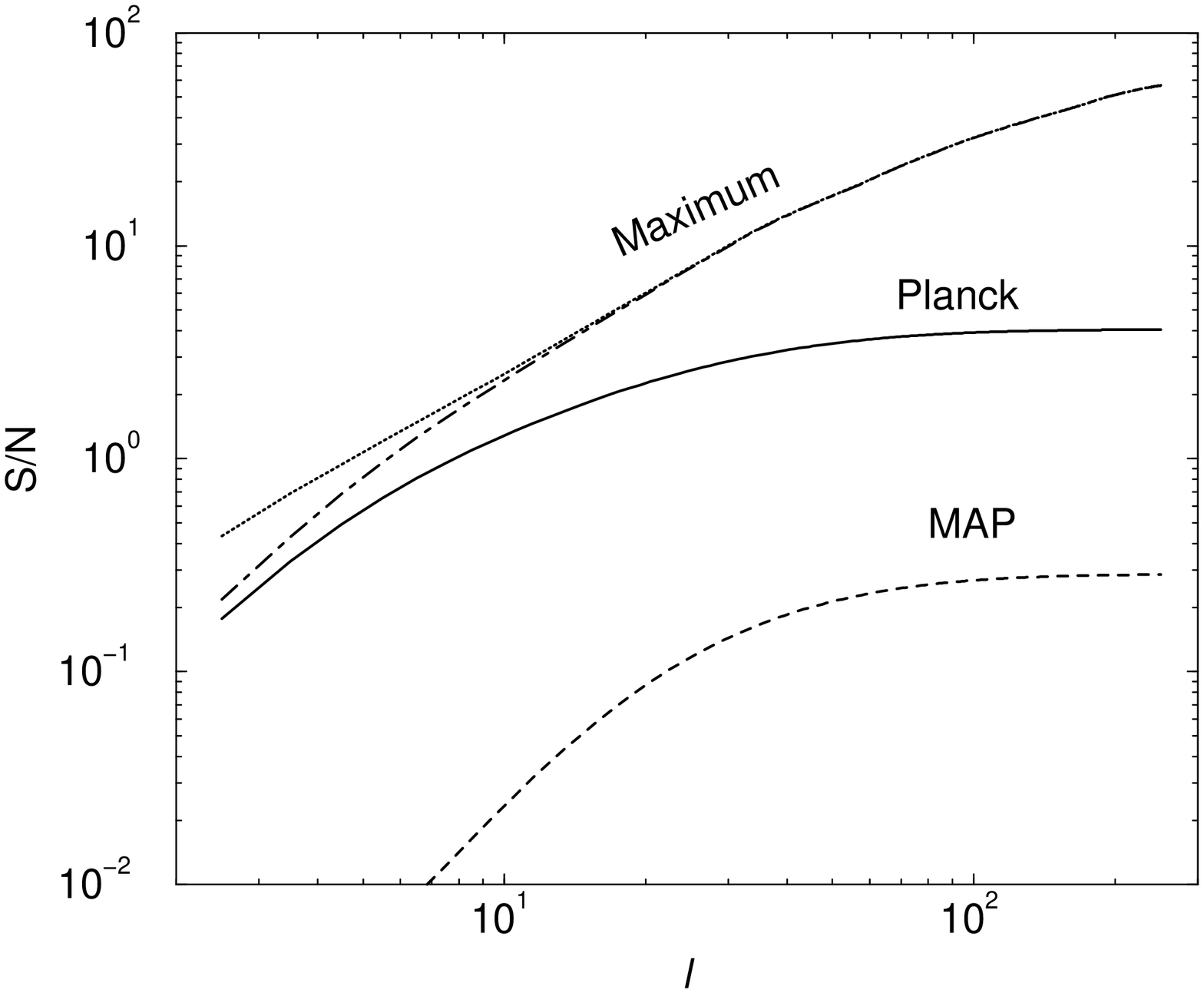,width=3.6in,angle=-90}}
\caption{{\it Left:} The cross-correlation power spectrum between the ISW and
SZ effects. For comparison, we also show the power spectra of
ISW and SZ effects and the CMB anisotropies. {\it Right:} The cumulative
signal-to-noise for the detection of the ISW-SZ correlation with Planck
and MAP data and using spectral dependence of the SZ contribution to
separate it out from thermal CMB fluctuations. The maximum signal-to-noise
is when the ISW effect is separated from the dominant primary
fluctuations at the last scattering surface.}
\label{fig:iswsz}
\end{figure*}

\subsection{ISW-SZ Correlation}

Following the derivation of the ISW-large scale structure correlation,
we can also consider the cross-correlation between the ISW effect and
the Sunyave-Zel'dovich thermal effect \cite{SunZel80} 
due to inverse-Compton scattering of CMB photons via hot electrons. 
The temperature decrement due to the SZ effect can be written as
the integral of pressure along the line of sight
\begin{equation}
y\equiv\frac{\Delta T}{T_{\rm CMB}} = g(x) \int  d\rad  a(\rad)
\frac{k_B
\sigma_T}{m_e c^2} n_e(\rad) T_e(\rad) \,
\end{equation}
where $\sigma_T$ is the Thomson cross-section, $n_e$ is the electron
number density, $\rad$ is the comoving distance, and $g(x)=x{\rm
coth}(x/2) -4$ with $x=h \nu/k_B
T_{\rm CMB}$ is the spectral shape of SZ effect. At Rayleigh-Jeans
(RJ) part of the CMB, $g(x)=-2$.
For the rest of this paper, we assume observations in the Rayleigh-Jeans
regime of the spectrum. Due to the spectral
dependence of the SZ effect when compared to CMB thermal fluctuations,
the SZ signal can be extracted from CMB fluctuations in
multifrequency data \cite{Cooetal00}.
Here, we use expected results from such a frequency separation 
with Planck and MAP data, 
and consider the ISW-SZ cross-correlation by correlating the CMB and SZ maps. 
When calculating signal-to-noise, we  
will use the noise power spectra calculated in \cite{Cooetal00}
for the Planck SZ and CMB maps.

Following our discussion on the ISW-large scale structure correlation,
we can write the relevant window function for the SZ effect as
\begin{equation}
W^{\rm SZ}(r) = g(x) \frac{k_B \sigma_T b_g(\rad) \bar{n}_e}{a(r)^2 m_e c^2}
\label{eqn:wsz}
\end{equation}
where, at linear scales corresponding to the ISW effect,
 the pressure bias, $b_g$, relative the density field follows
from arguments based on the halo approach to large scale pressure
fluctuations \cite{Sel00,Coo01}:
\begin{equation}
b_g(z) = \int dM \frac{M}{\bar{\rho}}\frac{dn(M,z)}{dM} b_{\rm halo}(M,z) 
T_e(M,z) \, .
\end{equation}
Here, $T_e(M,z)$ is the electron temperature, which can be
calculated through arguments related to the virial theorem.
In Eq.~\ref{eqn:wsz}, $\bar{n}_e$ is the mean density of electrons
today. 

In figure~\ref{fig:iswsz}, we show the cross-correlation between
the ISW and SZ effects. For comparison, we also show the ISW and SZ power
spectra. The correlation coefficient for the ISW-SZ effect ranges
from 0.3, at $l \sim$ few tens to 0.1 at $l \sim$ few hundred suggesting
that ISW and the SZ contributions are not strongly correlated.
This could be understood based on the fact that contributions to the
SZ effect primarily comes  from the so-called 1-halo term of the 
halo models of non-linear clustering and not the 2-halo term that
tracers the large scale correlations and, thus, the linear density fluctuations
responsible for the ISW effect. 

In the same figure, we also show
the signal-to-noise for the detection of the ISW-SZ cross-correlation.
For an experiment like Planck, we find that the signal-to-noise
ratio is at the level of $\sim$ 4, while for MAP, it is at the level of
0.3; this is understandable as MAP has no high frequency information for
a reliable separation of the SZ effect. In order to investigate what
limits the signal-to-noise for the detection of the ISW-SZ correlation
with a mission like Planck, we decided to set the noise contribution
to the ISW effect as simply due to the ISW effect itself, instead of
the total CMB power spectrum. This led to the dot-dashed line.
Further removing the noise contribution to the SZ map, such that a perfect
separation of the SZ effect is possible led to the  dotted line, which
converges to the previous dot-dashed line by $l \sim 10$. It is
clear that the dominant noise contribution comes from the 
noise associated with the ISW effect, and effectively, the dominant
CMB anisotropies at the last scattering
which cannot be easily separated from the ISW effect.
If there is a useful separation scheme to extract the ISW contribution alone,
the expected signal-to-noise for the ISW-SZ correlation is
at the level of $\sim$  60 suggesting a clear detection of the
correlation which can be used for cosmological and astrophysical purposes.

\begin{figure}
\centerline{\psfig{file=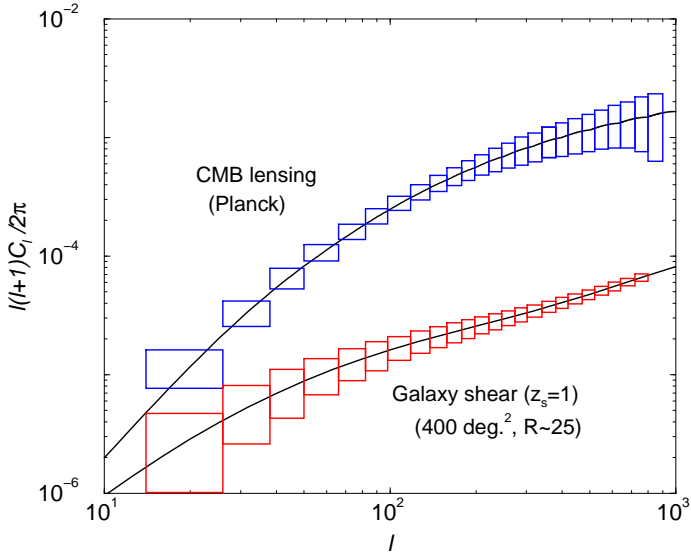,width=3.6in,angle=-90}}
\caption{The power spectrum of convergence constructed from
CMB deflections (top curve) and galaxy
shape data (bottom curve). In the case of reconstruction based
on CMB, we show expected errors from the Planck mission
while for large scale structure weak lensing, we
show expected errors for a survey of 400 deg.$^2$ down to a R band
magnitude of 25.}
\label{fig:lensing}
\end{figure}

\subsection{ISW-lensing correlation}

We can consider two forms of lensing: the potentials that deflect
CMB photons and potentials that shear background galaxy images.
The former can be constructed using quadratic statistics in temperature data
or using Fourier space statistics that are optimized to
extract the lensing signal, while the latter is probed in
weak lensing survey using galaxy shapes. 

The deflection angle of CMB photons on the sky,
$\alpha(\bn) = \nabla \phi(\bn)$, are given by the
gradient of the projected potential $\Phi$ (see e.g. \cite{Kai92}),
\begin{eqnarray}
\phi(\bm)&=&- 2 \int_0^{\rad_0} d\rad
\frac{\da(\rad_0-\rad)}{\da(\rad)\da(\rad_0)}
\Phi (\rad,\hat{{\bf m}}\rad ) \,.
\label{eqn:lenspotential}
\end{eqnarray}
The lensing potential can be related to the well known convergence generally
 encountered in conventional lensing
studies involving galaxy shear 
\begin{eqnarray}
\kappa(\bm) & =& {1 \over 2} \nabla^2 \phi(\bm) \nonumber \\
& = &-\int_0^{\rad_0} d\rad \frac{\da(\rad)\da(\rad_0-\rad)}{\da(\rad_0)}
\nabla_{\perp}^2 \Phi (\rad ,\hat{{\bf m}}\rad) \, , \nonumber \\
\end{eqnarray}
where note that the 2D Laplacian operating on
$\Phi$ is a spatial and not an angular Laplacian.
Expanding the lensing potential to Fourier moments,
\begin{equation}
\phi(\bn) = \int \frac{d^2\vecl}{(2\pi)^2} \phi(\vecl)
{\rm e}^{i \vecl \cdot \bn}  \, ,
\end{equation}
we can write the usually familiar quantities of convergence and shear 
components of weak lensing as \cite{Hu00}
\begin{eqnarray}
\kappa(\bn) &=& -\frac{1}{2}\int \frac{d^2\vecl}{(2\pi)^2}
l^2 \phi(\vecl) {\rm e}^{i\vecl \cdot \bn} \nonumber \\
\gamma_1(\bn) \pm i\gamma_2(\bn) &=& -\frac{1}{2}\int \frac{d^2\vecl}{(2\pi)^2}\ 
l^2 \phi(\vecl) {\rm e}^{\pm i 2 (\phi_l-\phi)}{\rm e}^{i\vecl \cdot \bn} \, .
\label{eqn:kappaphi}
\end{eqnarray}
Though the two terms $\kappa$ and $\phi$ contain
differences with respect to radial and wavenumber weights,
these differences cancel with the Limber approximation \cite{Lim54}.      
In particular, their spherical harmonic moments are simply proportional
\begin{eqnarray}
\phi_{l m} &=&
             -{2 \over l(l+1)} \kappa_{l m} =
                 \int d {\bn} \Ylmn{}^*(\bn) \phi(\bn) \nonumber\\
             &=& i^l \int {d^3 {\bf k}\over 2\pi^2} \delta({\bf k})
                \Ylmn{}^* (\bk) I_\ell^{\rm len}(k)
\label{eqn:GSSZequiv}
\end{eqnarray}
with
\begin{eqnarray}
I_\ell^{\rm len}(k)& =& \int W^{\rm len}(k,r)
                 j_l(k\rad)  \,,\nonumber\\
W^{\rm len}(k,r)& =&
                -3 \frac{\Omega_m}{a} \left({H_0 \over k}\right)^2
                G(r) {\da(\rad_0 - \rad) \over
                \da(\rad)\da(\rad_0)}\,.
\label{eqn:lensint}
\end{eqnarray}
Here, we have used the Rayleigh expansion of a plane wave, 
equation~(\ref{eqn:Rayleigh}),
and the fact that $\nabla^2 \Ylmn = -l(l+1) \Ylmn$.

\begin{figure}
\centerline{\psfig{file=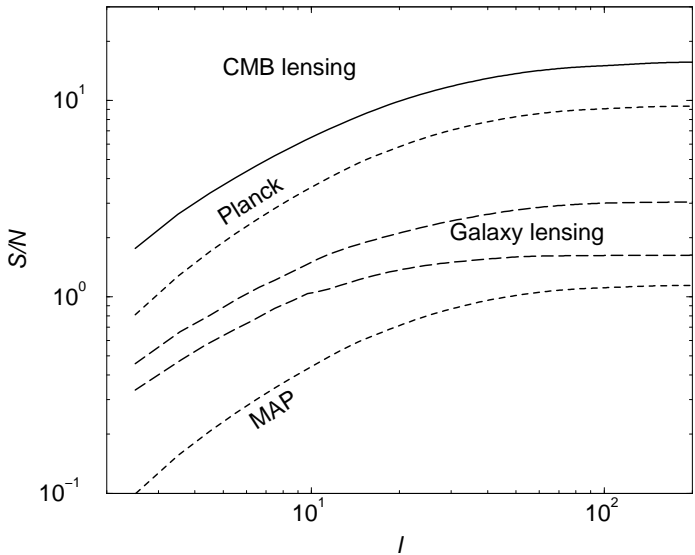,width=3.6in,angle=-90}}
\caption{The signal-to-noise for the detection of lensing-ISW
cross-correlation with galaxy shear data and using deflections in
the CMB. The dotted lines are for galaxy weak lensing surveys with
background sources at redshifts of 0.5 and 1.5, respectively, and with
a sky area of $\pi$ steradians, similar to the Sloan survey. 
The dashed lines show the signal-to-noise when 
temperature data is cross-correlated with an estimator
for lensing deflections in Planck and MAP temperature data. The solid line
is when a temperature map is cross-correlated with
a noise-free estimator of deflections.}
\label{fig:lensingsn}
\end{figure}

For the construction of deflection angles based on the CMB temperature data,
we make use the quadratic statistic proposed by \cite{Hu01} involving
the divergence of the temperature weighted temperature gradients, $\nabla
\cdot (\theta \nabla \theta)$. In Fourier space, we can write the
estimator for the deflection angles as
\begin{eqnarray}
D(\vecl) &=& \frac{N_l}{l} \int \frac{d^2\vecl_1}{(2\pi)^2} \left(\vecl \cdot
\vecl_1 C_{l_1}^{\rm CMB} +\vecl \cdot (\vecl-\vecl_1) C_{|\vecl-\vecl_1|}^{\rm CMB}\right)\nonumber \\
&& \quad \times \frac{\theta(l_1)\theta(|\vecl-\vecl_1|)}{2 C_{l_1}^{\rm tot}
C_{|\vecl-\vecl_1|}^{\rm tot}} \, .
\end{eqnarray}
The ensemble average, $\langle D(\vecl) \rangle$, is equal to
the deflection angle, $l\phi(\vecl)$, when
\begin{equation}
N_l^{-1} = \frac{1}{l^2} 
\int \frac{d^2\vecl_1}{(2\pi)^2} \frac{\left(\vecl \cdot
\vecl_1 C_{l_1}^{\rm CMB} +\vecl \cdot (\vecl-\vecl_1) C_{|\vecl-\vecl_1|}^{\rm CMB}\right)^2}{2 C_{l_1}^{\rm tot}
C_{|\vecl-\vecl_1|}^{\rm tot}} \, .
\end{equation}
Note that $N_l$ is the noise power spectrum associated with the
reconstructed deflection angle power spectrum:
\begin{equation}
\langle D(\vecl) D(\vecl') \rangle = (2\pi)^2 \delta_D(\vecl+\vecl') \left(l^2 C_l^{\phi \phi} +N_l\right) \, .
\end{equation}

In the case of lensing surveys using galaxy shear data,
we rewrite equations~(\ref{eqn:kappaphi})
and (\ref{eqn:lensint}) such that $\da(\rad_0) = \da(\rad_s)$
where $\rad_s)$ is radial distance to background sources from which
shape measurements are made. We assume that all sources are at
the same redshift, though,a distribution of sources in the redshift
range expected does not lead to a significantly different result than the
one suggested here. 

The shot-noise contribution to the convergence power spectrum associated 
with lensing surveys involving galaxy ellipticity data is
\begin{equation}
C_l^{\rm N_\X} = \frac{\langle \gamma_{\rm int}^2 \rangle}{\bar{n}} \, ,
\end{equation}
where $\langle \gamma_{\rm int}^2 \rangle^{1/2}$ is the rms noise
per component introduced by intrinsic ellipticities, typically $\sim$ 0.6
for best ground based surveys, and $\bar{n}$ is the surface number density
of background source galaxies from which shape measurements can
be made. For surveys that reach a limiting magnitude in $R \sim 25$,
the surface density is consistent with $\bar{n} \sim 6.9 \times 10^8$ sr$^{-1}$
or $\approx 56$ gal arcmin$^{-2}$ \cite{Smaetal95}, 
such that $C_l^{\rm N} \sim 2.3 \times
10^{-10}$.

In figure~\ref{fig:lensing}, we compare the lensing convergence
power spectrum associated with CMB (top curve) and a large scale
structure weak lensing survey from galaxy ellipticities with
background source galaxies at a redshift of one.
Note that we have obtained the convergence power spectrum associated with
lensing deflections in CMB following
the estimator for the lensing
deflection power spectrum and the two are simply related
following equation~(\ref{eqn:kappaphi}) such that
$C_l^{\kappa} = l^4/4 C_l^{\phi\phi}$.
For comparison, we also show expected error bars on the reconstructed
convergence power spectrum from CMB, via the Planck temperature data,
and for a wide-field survey of 400 deg.$^2$ down to a R-band magnitude of
25.

In figure~\ref{fig:lensingsn}, we show the associated cumulative
signal-to-noise in the detection of the ISW-lensing correlation
for galaxy lensing surveys and using CMB data. 
We assume an area of $\pi$ steradians for the lensing surveys
and as the signal-to-noise scales as $f_{\rm sky}^{1/2}$, we do not
expect a significant use of the current and upcoming lensing
surveys which are restricted to at most few hundred sqr. degrees.
The dedicated instruments, such as the
Large-aperture Synoptic Survey Telescope (LSST; \cite{TysAng00}) 
however, will provide wide-area maps of the lensing convergence and
these will certainly be useful for cross-correlation studies with CMB to
extract the ISW effect.

Note that the Planck data allow the best opportunity to detect the ISW effect
by correlating an estimator for deflections with a temperature map.
The MAP has a lower cumulative signal-to-noise due to its
estimator of deflections is affected by the low resolution of the
temperature data. Nevertheless, the MAP data will certainly allow
the first opportunity to detect the presence of the ISW effect 
either from CMB data alone or through cross-correlation of
other tracers.

\begin{figure}
\centerline{\psfig{file=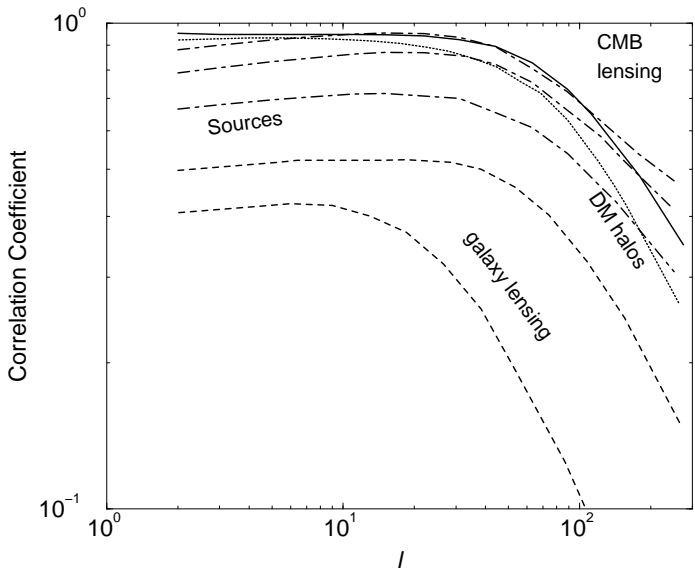,width=3.6in,angle=-90}}
\caption{The correlation-coefficient for ISW-large scale
structure correlations involving lensing effect on CMB (solid line),
a catalog of dark matter halos down to a mass limit of $10^{14}$ M$_{\sun}$
at all redshifts (dotted line), lensing convergence from galaxy shear data
(dashed lines) with sources at redshifts of 0.5 and 1.5, and
sources as tracers with mean redshifts of 0.4, 0.7 and 1.3. The
potentials that deflect CMB photons and sources at
$z \sim 1.5$ are best correlated with the ISW effect.}
\label{fig:corr}
\end{figure}

\begin{figure}
\centerline{\psfig{file=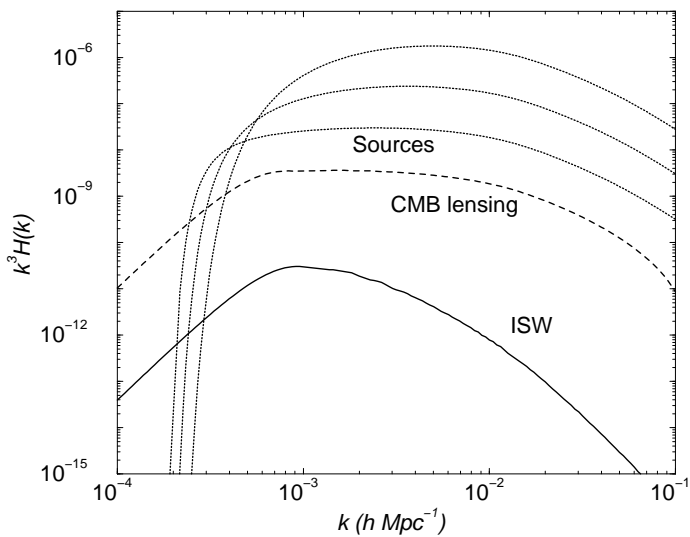,width=3.6in,angle=-90}}
\caption{The contribution to the power spectra of ISW effect and
other tracers as indicated. 
The sources have a mean redshift, $z_0$ of 0.4, 0.7 and 1.3 (from
top to bottom). The plot shows that the contribution
to the lensing effect on CMB comes from same Fourier modes as the
ISW effect while there is a  mismatch when compared to
galaxy surveys at low redshifts.} 
\label{fig:kspace}
\end{figure}

\section{Summary}

We discussed the correlation between late-time integrated Sachs-Wolfe (ISW)
effect in the cosmic microwave background (CMB) temperature anisotropies and
the large scale structure of the local universe. 
This correlation has been proposed  
and studied in the literature as a probe of the dark energy and its
physical properties. We considered a variety of large scale structure tracers 
suitable for a detection of the ISW effect via a 
cross-correlation. 

We summarize our results for the correlation coefficient
in figure~\ref{fig:corr}. As shown, the potentials that deflect CMB photons
are strongly correlated with the ISW effect with correlation coefficients
of order 0.95 at multipoles of $\sim$ 10. The convergence maps constructed
from galaxy shear data are not strongly correlated due to mismatches 
in their window  function and that of the ISW effect. While catalogs of
dark matter halos, or galaxy clusters, are well correlated with the
ISW effect, sources at low redshifts necessarily not. Increasing the
mean source redshift to $\sim$ 1.5, we find correlation coefficients of
$\sim$ 0.9 suggesting that a catalog of sources at such high redshifts are
preferred over a low redshift survey. Such catalogs can be constructed
from surveys at X-ray and radio wavelengths, however, 
the surface density of such sources are significant lower than 
optical galaxies and the resulting cumulative signal-to-noise
is subsequently smaller.

 In order to understand the correlation between ISW effect and 
tracers, we rewrite equation~(\ref{eqn:clexact}) in the
form $\int dk/k k^3 H(k)$. In figure~\ref{fig:kspace}, we
show $k^3H(k)$ as a function of $k$ for large scale structure
tracers involving sources and the ISW effect. For comparison,
we also show the contribution to potential  power
spectrum that deflect CMB photons. As shown, there is a
mismatch between the wave numbers that contribute to
the galaxy clustering, at low redshifts,
 when compared to the ISW effect. With increase in source redshift,
the mismatch decreases and the sources tend to be more correlated with
the ISW effect. On the other hand,
it is clear that lensing potentials that deflect CMB
photons are well correlated with the ISW effect. 
In fact, we find that the Planck survey data alone provide the best
opportunity to extract the ISW effect using an estimator
of lensing deflections on its temperature data.
With multifrequency data, the presence
of the ISW-large scale structure correlation can also be investigated through
a cross-correlation of the frequency cleaned SZ and CMB maps.

In the near term, the ISW-tracer correlation is best studied through
catalogs of clusters selected based on mass.
Such catalogs  can be constructed with upcoming wide-field lensing, 
SZ effect and X-ray imaging surveys. 
The combined DUET cluster catalog
and the MAP data will allow a good opportunity to study the 
ISW-large scale structure correlation  while the MAP data can also
be used with the Sloan catalog of galaxies. We estimate signal-to-noise of
order few for these scenarios suggesting that any such detection will
be challenging. For DUET, attempts to improve the shot-noise on 
the X-ray side will be necessary while for Sloan, one should carefully
select a subsample of high redshift galaxies, say through photometric
redshift data. 

\acknowledgments
This work was supported at Caltech by the Sherman Fairchild foundation and the
DOE grant DE-FG03-92-ER40701.

\end{document}